\documentclass[AMA,LATO1COL]{WileyNJD-v2}

\articletype{Article}%

\received{26 April 2016}
\revised{6 June 2016}
\accepted{6 June 2016}

\usepackage{amsmath}
\usepackage{subfig}

\usepackage{ifthen}
\usepackage{amssymb}
\usepackage{xcolor}
\usepackage{ulem}
\usepackage{balance}

\newboolean{showcomments}
\setboolean{showcomments}{true}
\ifthenelse{\boolean{showcomments}}
	{\newcommand{\nb}[3]{
		{\colorbox{#2}{\bfseries\sffamily\scriptsize\textcolor{white}{#1}}}
		{\textcolor{#2}{\sf\small$\blacktriangleright$\textit{#3}$\blacktriangleleft$}}}
	 }
	{\newcommand{\nb}[3]{}
	 }

\begin{document}
\title{%An Approach for 
Retrieving and Ranking Relevant JavaScript Technologies from Web Repositories}
%% Por ahi la ultima parte de "using..." se puede sacar

%% OTRA ALTERNATIVA MAS PROVOCADORA
%% "Tackling Technology Fatigue in Javascript: An Approach for Recovering Relevant Technologies from Web Repositories"

\author[1]{Hernan C. Vazquez}

\author[1]{J. Andres Diaz Pace}

\author[2]{Claudia Marcos}

\author[1]{Santiago Vidal*}

\authormark{H.C. Vazquez \textsc{et al}}

\address[1]{\orgname{ISISTAN-CONICET}, \orgaddress{\state{Tandil, Buenos Aires}, \country{Argentina}}}

\address[2]{\orgname{ISISTAN-CIC}, \orgaddress{\state{Tandil, Buenos Aires}, \country{Argentina}}}

\corres{Santiago Vidal, ISISTAN-CONICET. \\ \email{svidal@exa.unicen.edu.ar}}

\abstract[Summary]{The selection of software technologies is an important but complex task. We consider developers of JavaScript (JS) applications, for whom the assessment of JS libraries has become difficult and time-consuming due to the growing number of technology options available. A common strategy is to browse software repositories via search engines (e.g., NPM, or Google), although it brings some problems. First, given a technology need, the engines might return a long list of results, which often causes information overload issues. Second, the results should be ranked according to criteria of interest for the developer. However, deciding how to weight these criteria to make a decision is not straightforward. In this work, we propose a two-phase approach for assisting developers to retrieve and rank JS technologies in a semi-automated fashion. The first-phase (ST-Retrieval) uses a meta-search technique for collecting JS technologies that meet the developer's needs. The second-phase (called ST-Rank), relies on a machine learning technique to infer, based on criteria used by other projects in the Web, a ranking of the output of ST-Retrieval. We evaluated our approach with NPM and obtained satisfactory results in terms of the accuracy of the technologies retrieved and the order in which they were ranked.}

%the lack of precision in the search engines available in web software repositories. Because of this problem, developers and architects often prefer general-purpose search engines, which hold promises for a better precision, but force decision-makers to navigate within each obtained result and lead to information overload issues. In this work, we propose a meta-search approach that automates the process of retrieving software technologies for a specific design or development need. This approach, called ST-Retrieval, is based on the extraction, filtering and aggregation of software technologies returned by various existing search engines. ST-Retrieval was evaluated in one of today's largest software repositories (the NPM repository for JavaScript) with promising results.

\jnlcitation{\cname{%
\author{H. Vazquez}, 
\author{J.A. Diaz Pace}, 
\author{C. Marcos}, and
\author{S. Vidal}} (\cyear{2020}), 
\ctitle{An Automated Machine Learning Approach for Retrieving and Ranking Relevant JavaScript Technologies from Web Repositories}, \cjournal{Software Evolution and Process}, \cvol{2017;00:1--6}.}

\maketitle

%\footnotetext{\textbf{Abbreviations:} ANA, anti-nuclear antibodies; APC, antigen-presenting cells; IRF, interferon regulatory factor}

\section{Introduction}
Software technologies are an essential part of current software development
practices. The use of software technologies, such as libraries and
frameworks, can greatly improve developers' productivity.
%, which permits to accelerate development times and deliver value to customers quickly
Nonetheless, an inappropriate technology selection can negatively affect the software product being built, and also the business goals of the organization \cite{lin2007fuzzy}.
In this context, the task of selecting a software technology that fulfills the specific needs of a development task is generally a complex and time-consuming decision-making process. 
%Furthermore, the selection task is often assigned under schedule pressures and developers may not have enough time or experience to plan the selection process in detail \cite{kontio1996case}. 
%Nonetheless, an inappropriate technology selection can negatively affect the software product being built, and also the business processes of the organization \cite{lin2007fuzzy}.
One of the reasons for this complexity is the availability of a large number of software technologies in the market, in response to the growing demand from software companies. Keeping up-to-date with technological developments is challenging for developers. 

In particular, this situation is very common in JavaScript (JS) development, as developers have to regularly search, evaluate and compare candidate JS libraries/frameworks for their applications. 
%and ii) the lack of technical knowledge and experience in the decision-makers.
%Thus, the process of evaluating and comparing candidate technologies
%ends up consuming a lot of time and being tedious %to those in charge of carrying out the task
This process can be perceived by developers as a ``technological fatigue"\footnote{https://medium.com/@ericclemmons/javascript-fatigue-48d4011b6fc4}.
%In this work, we focus on the case of JavaScript (JS), which
%is one of the most popular languages today \footnote{http://redmonk.com/sogrady/2017/03/17/language-rankings-1-17/}.
This phenomenon is due to the extensive number of JS technologies available in Web repositories, such as NPM\footnote{https://www.npmjs.com/} (Node Package Manager), which promote reuse by lowering production
costs and speeding up software delivery \cite{wittern2016look}. %Many packages are added to
%JS software repositories and many other packages get updated 

%In this context, the selection of JS technologies becomes a complex problem for developers, who have begun to report ``technological fatigue" when picking JS technologies for their applications. 
We argue that one of the reasons for JS technological fatigue is the lack
of precision in the search engines for JS repositories. As a consequence, developers often resort to general-purpose search engines (e.g.,
Google or Bing) with the hope of having better results.
However, the downside of such engines is that they tend to return long lists of documents, and then developers have to navigate within each result to find possible JS technologies. This leads to information overload issues. %Furthermore, many of these results generally correspond to questions-answers pages (e.g. StackOverflow, or Quora, among others), which in turn have links to other web pages. 
Furthermore, once the developer identifies a set of candidate technologies for her application, she must analyzed each technology to decide which one better fits her needs. Normally, this decision is driven by features of the technology, such as: popularity in the community, or number of downloads, among others. Assigning weights to these features for comparison purposes is not straightforward. For instance, NPM uses the AHP (Analytic Hierarchy Process) \cite{saaty2008decision} technique to support comparisons of JS technologies.
%The whole search and retrieval process leads then to information overload issues. 

The problem of information overload has been studied in various disciplines \cite{eppler2004concept}, and Web search engines are one of
the main tools to face the problem \cite{berghel1997cyberspace}. The usage of search engines for Web-based software repositories has received some attention in the literature, but there are still challenges for finding technologies being relevant to a particular development (or technological) need \cite{Clayton}.
%In particular, the selection of technologies is a process that benefits directly from the search engines, since they help %architects and developers to find candidate technologies to solve their needs. Although search engines for web-based software repositories have received some attention in the literature, there are still several challenges for finding technologies relevant to a particular development activity or project \cite{Clayton}. 
Previous works on software technology
selection have addressed the problem from different perspectives \cite{birk1997modelling}\cite{basili1991support}\cite{klein2015design}\cite{grande2014framework}\cite{vegas2005characterisation}.
Nonetheless, most works focus on the
evaluation of candidate technologies based on specific characteristics,
departing from a given group of technologies, and they leave
aside the problem of searching for relevant candidates.  In addition, the characteristics for decision making are often manually assessed by developers. %\adp{Esta bueno un mini-related work aca, pero sin abusar. Por otro lado, estaria cubriendo solo lo de Retrieval, pero NO lo de Ranking}%%Some recent works \cite{ouni2017search}\cite{zhan2016dolphin} have proposed the creation of dedicated search engines with their own indexing and search mechanisms, but they do not leverage on the (often powerful) capabilities of current search engines.

In this article, we propose an approach to assist developers in searching and ranking JS technologies. The approach works in two phases called \emph{ST-Retrieval} and \emph{ST-Rank}.
Given a developer's query expressing a JS technological need, \emph{ST-Retrieval} applies a meta-search strategy \cite{aslam2001models}, which combines
the search results of both a JS-specific engine and general-purpose engines. 
%(e.g. NPM\footnote{https://www.npmjs.com/}) and general-purpose (e.g. Google) engines. %To this end, we formally describe the software technology retrieval problem and model it as a meta-search problem. Additionally, we propose a technique based on \emph{string-matching} \cite{navarro2001guided} for the extraction of technologies being mentioned in web documents. This technique permits a seamlessly inclusion of general-purpose search engines in the process of meta-searching software technologies. The inclusion of general-purpose search engines allows us to emulate the standard developers' search behavior and helps to improve the performance of repository search engines.
Based on the technologies recovered by \emph{ST-Retrieval},  \emph{ST-Rank} generates a ranking of those technologies for the developer by means of a pair-wise learning to rank method. The ranking is based on an (automated) analysis of technology features extracted from JS projects on the Web (e.g.,  number of stars in the repository,  number  of  releases  in  the  last  year, number of contributors, etc.). To assess the relevance of the projects, we employ a popularity metric called CDSel. CDSel measures the relationship between the number of projects in which the technology was selected and how popular are those projects.% (Community Degree of Selection).

We have evaluated the approach on the NPM repository for JS, using a predefined set of queries and 1000 popular projects from GitHub. For \emph{ST-Retrieval}, our experiments reported an average precision improvement of 20\%, and we were able to recover a larger number of relevant JS technologies than using the default search engine provided by NPM. Regarding \emph{ST-Rank}, we observed improvements of at least 20\% on average when compared to the default ranking strategy followed by NPM. Based on these initial results, our approach makes two contributions: (i) it can boost the performance of the NPM search
system by leveraging on results provided by multiple search
engines, and (ii) it helps developers by ranking first those technologies being widely used in the JS community. %Furthermore, we believe these are important advantages of \emph{ST-Retrieval} to support decision-makers in having better-informed options during their technology evaluations.

The rest of the article is structured as follows. In Section \ref{sec:STSP}, we provide a brief description of the software technology selection problem, along with a motivational scenario in JS development. In Section \ref{sec:approach}, we describe the two phases of  the approach in detail. Section \ref{sec:evaluation} presents the evaluation of \emph{ST-Retrieval} and \emph{ST-Rank}, and discusses their pros and cons. Section \ref{sec:relatedWork} covers related work. Finally, Section \ref{sec:Conclusiones} gives the conclusions and outlines future lines of work.

\section{Software Technology Selection}\label{sec:STSP}

%The selection of software technologies involves understanding and deciding what technologies (from a given collection) are best suited to solve a specific need \cite{grande2014framework}. Figure \ref{fig:Software-Technology-Selection} shows an overview of the software technology selection process. The process starts when a user (a developer or an architect) identifies a need for performing a given task, and this need could be solved through the use of an existing software technology (e.g., frameworks, libraries, packages, etc.). Often times, the need is expressed as a query. The next step is to search and retrieve a sub-set of candidate technologies that might solve the problem from a larger set of available technologies. Then, according to the context of the project or the target task, the candidate technologies are assessed and compared in terms of their pros and cons so as to make a final selection. At last, if the user is satisfied with the selected technology, she will proceed to apply it. If not, she might perform another search with a different query. This process starts again as different needs arise.

%\begin{figure}
%	\begin{centering}
%		\includegraphics[scale=0.68]{retrieval_files/STSP2}
%		\par\end{centering}
%	\caption{Software Technology Selection Overview. \label{fig:Software-Technology-Selection}}
	
%\end{figure}

From a development perspective, the selection of software
technologies has an influential role in both
the development process and the quality of the final product \cite{jadhav2011framework}.
The successful application of a given technology (e.g., a JS
package) means that its usage for a particular task produces a desired objective \cite{birk1997modelling}. This success also depends on contextual features (e.g., alignment between the developer's requirement and the chosen package, maintenance of the package, type of license, or package usability, among others). 

As a motivational example of the technology selection problem for JS development, let us consider a JS developer that
needs to extract a barcode from an image, with the goal of automating
a process for extracting codes from a series of image files. The application context for this functionality is a Web browser. %\adp{Necesitamos motivar explicitamente el application context en el ejemplo, para que despues sea "natural" plantearlo en el approach!}\sv{Me parece que aca no porque ensucia el ejemplo. Si yo digo que quiero una tecnologia para node.js ¿porque no agregar Node.js al query?}%(Figure \ref{fig:Manual-retrieval-process.}).
%\begin{figure*}
%	\begin{centering}
%		\includegraphics[scale=0.6]{retrieval_files/motivexamp}
%		\par\end{centering}
%	\caption{Motivational example of searching a software technology. \label{fig:Manual-retrieval-process.}}
%\end{figure*}
Initially, she goes to the NPM package repository and submits the query ``extract barcode from image'' to its search engine, which returns only the package \emph{bytescout}\footnote{https://bytescout.com/}
as output. Bytescout is a JS client for a cloud service of the same
name. When reading about Bytescout, our developer finds out that it
is a paid service and that the JS client is not open-source. Also,
when she looks at the package description, NPM reports that Bytescout has been downloaded 40 times in the last month, which for a JS package might indicate that it is not very popular in the community. Let us assume that our developer is not convinced with these features, or
they are not aligned with the standards of her project. However, bytescout is the only technology returned by NPM. 
In this context, our developer has several options, namely: (i) adopt the package
despite her disagreement with its features, (ii) implement a
solution from scratch to read the barcodes, (iii) try a modified query
with the hope of getting more results from NPM, or (iv) use alternative
information sources (e.g. Google, or NPMSearch, among other engines)
to find alternative technologies. Let us suppose here that she goes for the third option and re-phrases the query as ``barcode
reader", which makes NPM return now 16 results. After inspecting each result, our developer is still unconvinced of using any of
the technologies, since they do not seem to be very
popular nor receive enough maintenance. The scenario so far shows the current limitations of JS-specific search engines, like NPM.

Next, let us assume that our developer decides to go for the fourth option, and she submits the query ``extract
barcode from image javascript package" to Google (the last phrase
intend to prevent Google form returning results for libraries in
other languages). This query returns a list of Web pages, and our
developer then inspects each page in order to check whether
some JS technologies are mentioned. In doing so, she realizes
that a technology called \emph{QuaggaJS}\footnote{https://serratus.github.io/quaggaJS/} is referenced in 3 results from the top-10 pages of the list. As she is not aware of this technology, she goes back to the NPM repository and
finds that QuaggaJS is more popular than bytescout, it
is open-source, and is well-maintained by the community. At this point,
our developer can pick QuaggaJS for her development need, or keep looking for other technologies. This scenario illustrates the challenge of using general-purpose search engines for JS, but also the issues related to the comparison of technologies.
%This situation exemplifies a number of typical questions faced by developers when selecting technologies: are there more candidate libraries to solve the problem?, am I selecting the correct library?, if I continue searching, will I find more useful technologies?
%, why does not NPM (as a JS-specific engine) return the result I found on Google (as a general-purpose engine)? Is there something wrong with the library proposed by Google?
%, among others.

%Based on the above, and also in our professional experience, 
The bottom line of the example has two implications. First, the search and comparison of JS technologies should take advantage of different information sources. Second, the
developer's manual analysis of technologies (e.g., by looking at Web sites) should be minimized, in the sense that semi-automated rankings could be created from criteria or feedback provided by other projects. Based on these ideas, we developed the \emph{ST-Retrieval} and \emph{ST-Rank} components of our approach.
%, particularly in cases like JS developments where the available technologies are hundreds of thousands
%. 

%In this context, we have developed  a two-phase approach for assisting developers to retrieve and rank JS technologies in a semi-automated fashion. 
%the \emph{ST-Retrieval} approach (Software Technologies Retrieval), whose main objective is the automation of the search and retrieval phases, based on the extraction and aggregation of the technologies collected by the search results of other existing engines.

\section{Approach}\label{sec:approach}
\begin{figure*}
	\begin{centering}
		\includegraphics[scale=0.68]{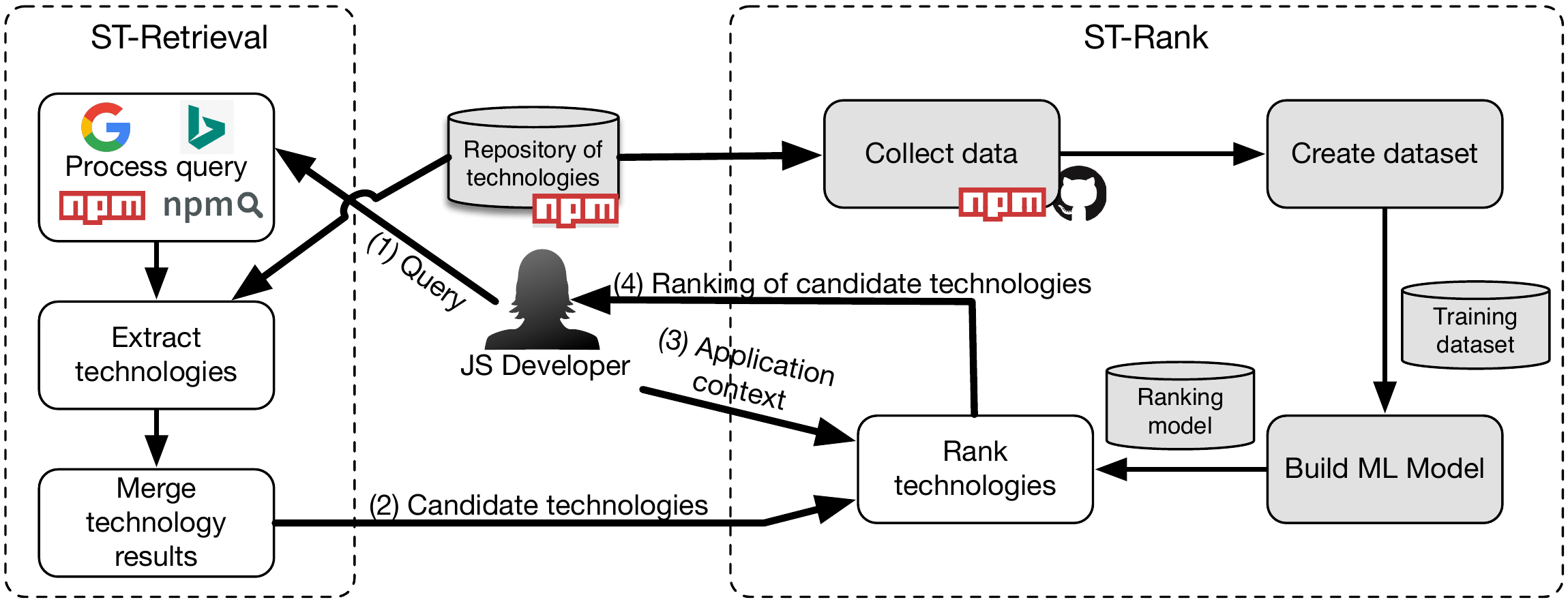}
		\par\end{centering}
	\caption{Overview of the two-phase approach. \label{fig:ApproachOverview}}
	
\end{figure*}

%In this section, we present our approach 
For retrieving and ranking relevant JS technologies, we propose two-phase approach as shown in Figure   \ref{fig:ApproachOverview}. The first phase, called \emph{ST-Retrieval}, takes a query given by the JS developer and returns a list of candidate technologies matching the query. The query is written in natural language and specifies a technological requirement (e.g. ``extract barcode from image"). The goal of this phase is to leverage on various search engines. Afterwards, the second phase, called \emph{ST-Rank}, is fed with the outputs of \emph{ST-Retrieval} along with an application context\footnote{Since some technologies only work in a specific context, our approach needs to be informed of the application context in which the developer is going to apply the technology} (i.e. Web browser, Node.js, etc.) provided by the developer, and generates a ranking of JS technologies based on their relevance. The goal of this phase is to infer a ranking by ``learning" from technology features and decisions made in other projects. To do so, a machine learning (ML) model is built. Each phase internally involves different steps. Gray boxes in the figure correspond to steps or artifacts provided at setup, while white boxes refer to steps performed every time a new query is entered by the developer.

%from repositories. Figure \ref{fig:ApproachOverview} shows the activities followed by the approach. The approach it is structured in two phases:  
%\begin{enumerate}
%    \item ST-Retrieval: the goal of this phase is to obtain a subset of JS technologies that match a query given by the developer.  The query is written is natural language by the developer and it specifies a technological requirement (e.g. ``extract barcode from image") 
%    \item ST-Rank: given the subset of JS technologies obtained in the previous phase and an application context (i.e. web browser, Node.js, etc.), this phase ranks the JS technologies according to their relevance using a machine learning algorithm. The algorithm is based on the analysis of the comparable features of each technology (e.g. keywords, number of releases in the last year, number of maintainers, etc.) and the decisions made by other projects.
%\end{enumerate}

%In the following subsections, we provide details of each phase of the approach and their inner activities. 

\subsection{ST-Retrieval}\label{sec:stretrieval}

We model the search and retrieval of JS technologies as a meta-search problem \cite{aslam2001models}.
In meta-search, the original query
is sent in parallel to a set of search engines, each one returning
an ordered list of items that satisfy the query.
The meta-search system combines all the lists into a new
one that is expected to keep the ``best" items of the individual lists.

The meta-search strategy for \emph{ST-Retrieval} works in three 
%helps developers to find JS technologies given a technological 
%need. \emph{ST-Retrieval} is composed by four 
steps (Figure \ref{fig:ApproachOverview}), as follows. First, the input query is sent to both domain-specific engines (e.g. NPM) and general-purpose engines (e.g., Google, Bing). The domain-specific engines directly return a list of JS technologies. However, the general-purpose engines produce a list of Web pages, which might include names of JS technologies. Second, these Web pages are processed in order to extract a list of JS technologies. Third, the lists from each search engine are merged into one single list, which is passed on to the second phase.

%namely:
%\begin{enumerate}
%    \item Process query: given a developer's query that expresses a technological need, this step uses different domain-specific (e.g. NPM) and general purpose (e.g. Google\footnote{https://www.google.com}) search engines to obtain a set of documents.
%\item Extract technologies: the documents retrieved in the previous %step are analyzed to extract the technologies present in them. his ocess varies according to kind of results returned by the search engine (e.g. NPM returns a list of JS packages, while Google returns a list of Web documents).
%    \item Create technology list by search engine: once the technology names were identified, this step creates a list of candidate technologies for each search engine. 
%    \item Merge technological lists: given the candidate lists of each search engine, this steps create a single candidate list that it is the output of \emph{ST-Retrieval}. 
%\end{enumerate}

The technical aspects of each step are explained below.

\subsubsection{Process query}\label{subsection:processQuery} In this step, we aim at leveraging on the capabilities of existing
search engines. %The main goal is to search for the query provided by the developer in different search engines. This search depends on the type of search engine under consideration. 
Specifically, we use four engines, namely: NPM,  NPMSearch\footnote{https://npmsearch.com}, Google,  and Bing\footnote{https://www.bing.com}. In general, the possible engines envisioned in our approach fall into two types: (i) \emph{domain-specific (DS)}, and (ii) \emph{general-purpose (GP)}. The former are especially designed
for the search of JS technologies (NPM and NPMSearch), while the latter are often used for general queries (Google and Bing).  %, and can be divided into two types: (a) \emph{Repository-Provided}  or (b) \emph{User-Created}. The difference between them is that the first type corresponds to the search engine provided by the de-facto repository (in our case NPM), while the second one is created by repository users in order to improve or facilitate their searches (in our case NPMSearch).

%(ii) \emph{General Purpose (GP)}: These search engines are commonly used
%for non-domain specific queries (Google and Bing).

Both types of engines allow developers to enter queries in natural language. Developers' queries can express functional or non-functional requirements that should be met by a JS technology. Stop-words (e.g., articles, pronouns, prepositions, etc.) that do not provide information to the search engine are removed from the query. For \emph{GP} engines, it  also augment the query to avoid results
from other domains (e.g., technologies in a language incompatible with
the project). Since we are retrieving JS technologies, the query is expanded with the suffix \char`\"{}javascript package\char`\"{}. Once the query was expanded, it is ready for execution in the corresponding engine. The result of this execution is a set of Web pages (or documents) in HTML, XML or JSON format.

\subsubsection{Extract technologies}\label{subsec:Technology-Extraction}
This step is concerned with obtaining the technologies named in the documents being returned by a query. Technologies are represented as a tuple $<name,\,repository_{-}url,\,home_{-}url,\,description>$
where:

\begin{itemize}
    \item \emph{name}: corresponds to the name of the technology
	in the repository (e.g., ``quagga'').
	\item \emph{repository\_url}:  is the technology URL on the repository
	site (e.g., ``https://www.npmjs.com/package/quagga'').
	\item \emph{home\_url}: is the URL of the technology site (e.g.,
	``https://serratus.github.io/quaggaJS/'').
	\item \emph{description}: is the description of the technology (e.g., ``An advanced barcode-scanner written in JavaScript'').
\end{itemize}

On one hand, for \emph{DS} engines, each result will map
one-to-one with existing technologies in the repository via its
\emph{name} and \emph{repository\_url}. On the other hand, for \emph{GP} engines, the search results are Web documents,
which might refer to zero, one or many technologies. For this reason,
those names of the technologies should be recovered. This information extraction task can be seen as a Named Entity Recognition (NER) problem. NER \cite{nadeau2007survey} aims at classifying entities found in a given text into predefined categories (e.g., people, organizations, places, time expressions, among others). In our work, the named entity category is \char`\"{}software
technology\char`\"{}, and we employ a rule-based strategy for string-matching
\cite{navarro2001guided}. By means of these rules, all the
technologies whose name or address (\emph{home\_url}, \emph{repository\_url})
match the category are extracted. 

\begin{figure}
	\centering{}{\includegraphics[scale=0.69]{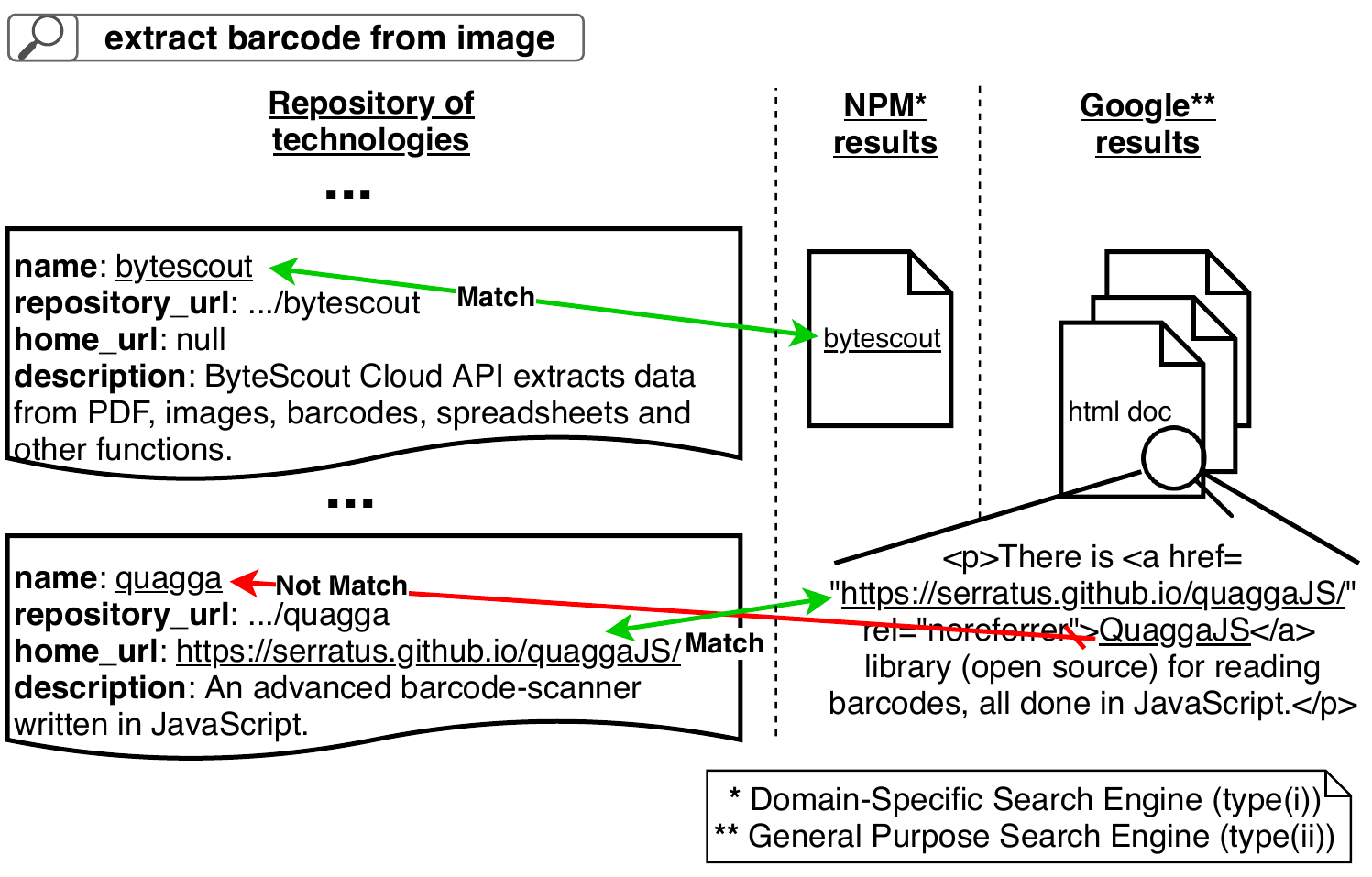}}\caption{Matching search results with technologies for different engines.\label{fig:Matching-Search-Results}}
\end{figure}

This step requires a repository of JS technologies as input (Figure  \ref{fig:ApproachOverview}),  which includes the information of the tuples. We generated this repository in advance via a process of Web crawling on the NPM site\footnote{In order to complement the information found in NPM, we only selected technologies whose repositories were publicly available in GitHub}. Figure  
\ref{fig:Matching-Search-Results} shows an example of results and  technologies obtained from the NPM and Google engines, along with
their mappings for the query ``extract barcode from image". In
the example, NPM returned a single result (bytescout) matching the name of a technology in the repository. Unlike NPM, Google returned an HTML document, whose text was parsed for matches of names or addresses of technologies. In particular, the  technology name in the resulting page (QuaggaJS) did not match the technology name in the repository
(quagga) but the address of the site did match (\emph{home\_url}, https://serratus.github.io/quaggaJS/).

\subsubsection{Merge technology results\label{subsec:Ranking-Creation}}
%The result of the previous step is a set of named technologies for each document. 
Based on the set of named technologies, this step creates an ordered list of technologies per search engine. For \emph{DS} engines, the sorting function is straightforward, and each technology is assigned
to the position of the result where it is named\footnote{this is because
each result corresponds unequivocally with a technology}. For \emph{GP} engines, a result can be related to many named technologies. In this case, the sorting function is based on the order in which the technologies were named. If two technologies are named within the same result, the first one named will appear in the list before than the one named afterwards. For example, if the first result contains the text \char`\"{}You can use Quagga or Barcode-Reader\char`\"{} and the next result contains the text \char`\"{}You should
use Bytescout\char`\"{}, then the candidates will be ordered as 
 1) Quagga, 2) Barcode- Reader, and 3) Bytescout. %\adp{Revisar ejemplo, no se puede hablar de "ranked document" yet, como decia antes}

%\begin{table}[h]
%	\small\sf\centering\caption{Differences between functions for different types of search engines.}
%		\begin{tabular}{lll}
%			\toprule
%			Function&Type(i)&Type(ii)\\
%			\midrule
%			Query&stop-words&stop-words + \\
%			Processing&&query expansion\\
%			\\
%			Technology& name matching & name matching +\\
%			Extraction&&url matching\\
%			\\
%			Ranking& result positions & result positions +\\
%			Creation&&order of mention\\
%			\bottomrule
%		\end{tabular}
%		NUNCA ES REFERENCIADA
%\end{table}

%\subsubsection{Merge technology lists}

After all search engines produced their individual lists, they are 
merged into a single list. In meta-search, this process is known
as a ranking aggregation function \cite{dwork2001rank}.
Aggregation functions can be classified into two main types \cite{liu2011learning}:
(i) those that use similarity scores returned by the search engine,
and (ii) those that use ranking positions. When engines do not
expose their similarity scores for their results, as in the
cases of Google or Bing, ranking positions can be used instead.
For this reason, we decided to use \emph{Borda Fuse} in our approach. Borda Fuse is a positional aggregation function that is computationally simple to implement \cite{dwork2001rank}, and it performs well in the context of Web searching \cite{aslam2001models}. 

In the Borda Fuse method,
%\emph{Borda Fuse} is an aggregation method based on the Borda Count choice strategy \cite{aslam2001models}. In this method, 
each search
engine is considered as a voter. Each voter presents a list of \emph{n}
ordered candidates (i.e., technologies). For each list, the best first candidate
receives \emph{n} points, the second candidate receives \emph{n-1}
points, and so on. Then, the points awarded by the different voters
are added and the candidates are ranked in descending order according
to the total of points obtained. Table \ref{tab:Borda-Fuse-Aggregation} shows an aggregation example using the \emph{Borda Fuse} method. Each list gives
a maximum of 4 points (equal to the maximum length of individual lists)
and decreases by one to each position of the list, i.e. the second
position gives 3 points, the third position gives 2 points, and the last
position gives 1 point. The scores are added and a final
list from highest to lowest scores is created. In our example, the most relevant results, quagga and bytescout, are in the top-ranked positions of the final list.

\begin{table}
	\caption{Borda Fuse aggregation example ({[}points{]} name). \label{tab:Borda-Fuse-Aggregation}}
	\begin{centering}
	\begin{tabular}{llll}
		\toprule 
		NPM & Google & Bing & Candidate list\\
		\midrule
		{[}4{]} \uline{bytescout} & {[}4{]} \uline{quagga} & {[}4{]} \uline{quagga} & {[}8{]} \uline{quagga}\\
		& {[}3{]} bcreader & {[}3{]} bc-js & {[}6{]} \uline{bytescout}\\
		& {[}2{]} \uline{bytescout} & {[}2{]} \emph{bwip-js} & {[}4{]} bcreader\\
		& {[}1{]} \emph{jaguar} & {[}1{]} bcreader & {[}3{]} bc-js\\
		&  &  & {[}2{]} \emph{bwip-js}\\
		&  &  & {[}1{]} \emph{jaguar}\\
		\bottomrule
	\end{tabular}
	\par \end{centering}
\end{table}

\subsection{ST-Rank}

\emph{ST-Retrieval} outputs a set of candidate technologies for the developer's query. Since these technologies have the same goals, the developer has to scrutinize them to find the technology that best fulfills her needs. This analysis often involves indicators provided by the repository (e.g., the number of downloads, dependent projects, or contributors, among others) and also searching for developers' opinions in blogs and forums. In this context, \emph{ST-Rank} assists developers to select the ``best" technology by creating a ranking based on the choices made by other developers. The assistance consists of four steps (Figure \ref{fig:ApproachOverview}), as follows. First, information about popular technologies used by other developers in open-source repositories is collected. Second, a dataset is created by taking into account the previous information and the application context of each technology (e.g., Web browser, Node.js, etc.). A training dataset is derived from the dataset, and we apply a supervised machine learning algorithm to build a ranking model for JS technologies. These three steps happen during the setup of the approach. At last, the machine learning model predicts a ranking for the technologies given by \emph{ST-Retrieval}, according to the ``patterns" inferred from the training dataset.

%\begin{itemize}
%    \item Collect data: this step collects information of the technological decisions made by developers in popular open-source repositories. Also, it collects features of the repositories analyzed (e.g. \#downloads, \#dependencies, etc.).
%    \item Create dataset: based on the information collected previously, this steps builds a training dataset that is going to be the input of a machine learning algorithm.  The dataset is built taking into account the application context of each technology (e.g.web browser, Node.js, etc.).
%    \item Build ML Model: a supervised machine algorithm is used to train a model to rank JS technologies.
%    \item Rank technologies: given a list of JS technologies and an application context, this steps rank the technologies based on the features. 
%\end{itemize}

The technical aspects of each step are explained below.

\begin{figure}
	\begin{centering}
	\subfloat[Collect data]{\includegraphics[scale=0.5]{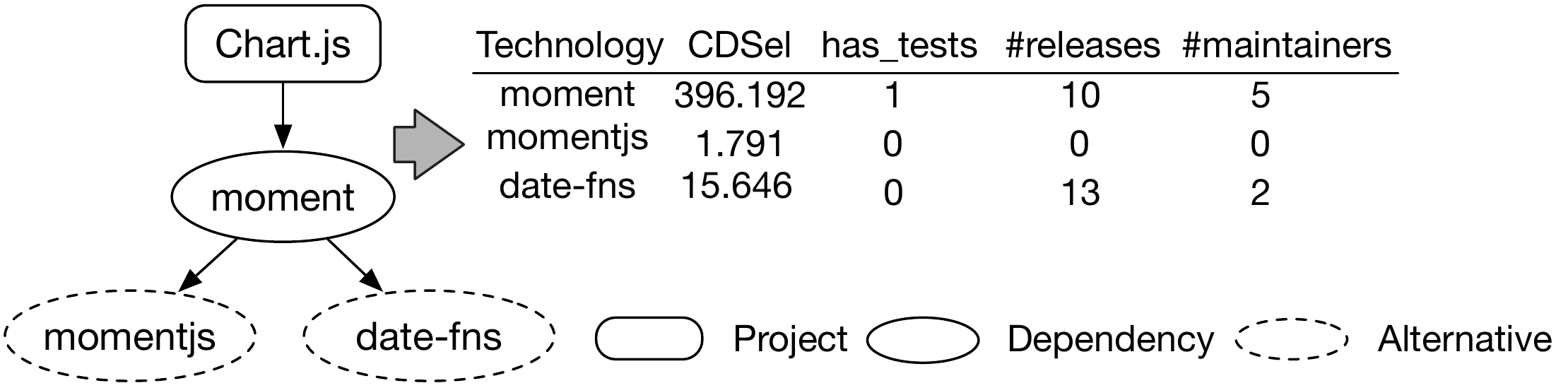}} \\
	\subfloat[Build dataset]{\includegraphics[scale=0.5]{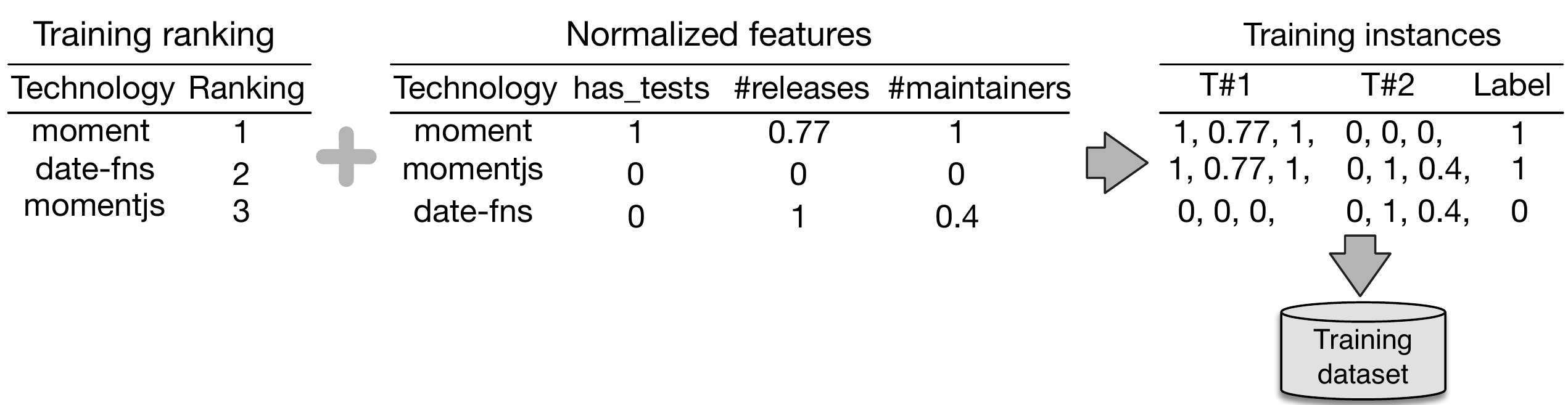}}
	\par \end{centering}
	\caption{Example of derivation of a training dataset.\label{fig:dataGeneration}}
\end{figure}

\subsubsection{Collect data}\label{subsec:collectData}
The input of this step is a group of popular open-source JS projects. We are interested in the dependencies of every project and their features. In NPM, these dependencies are in the \emph{package.json} of each project\footnote{Since both the ``reference projects" and their dependencies are JS projects themselves, for the sake of clarity we refer to the former as ``projects" and to the latter as ``technologies"}. 

The processing of each dependency involves three tasks. The first task is to look for alternatives to the technologies identified. For example, project \emph{Chart.js}\footnote{https://www.npmjs.com/package/chart.js} depends on \emph{moment}\footnote{https://www.npmjs.com/package/moment}, which is a library to manipulate dates (Figure \ref{fig:dataGeneration}). Thus, alternative technologies solving the same need are sought, such as \emph{date-fns}\footnote{https://www.npmjs.com/package/date-fns} and \emph{momentjs}\footnote{https://www.npmjs.com/package/momentjs}. We implemented this search by automatically scraping the website of NPMCompare\footnote{https://npmcompare.com}. For a given technology, the website gives a list of related packages.

The second task is to assess the ``popularity" of the dependency (or JS technology). To do so, we propose a metric called \emph{CDSel (Community Degree of Selection)} for a technology. CDSel is intended to capture the relationship between the number of projects in which the technology was selected and the relevance of those projects. Let $n$ be the number of reference projects, $rel(P_i)$ the relevance of project $i$, and $s=1$ when the technology was selected in $P_i$ (0 otherwise), then CDSel is computed as follows:
\[
CDSel=\sum_{i=1}^{n}s*\frac{rel(P_{i})}{log_{2}(i+1)}
\]

The logarithm serves as an attenuating factor in the ranking, so as to produce a controlled reduction in the values of the metric. For example, in our dataset we obtained a CDSel value of 396.192 for \emph{moment}, 15.646 for \emph{date-fns}, and 1.791 for \emph{momentjs}, which means that \emph{moment} is most often selected by relevant repositories than \emph{date-fns} and \emph{momentjs} (Figure \ref{fig:dataGeneration}). Regarding $rel(P_i)$, it is computed as follows:
\[
rel(P)_{i}=z-(rankPosition(P_{i}))
\]
where $z$ is the length of the ranking and $rankPosition(P_i)$ is the position in the ranking of project $(P_i)$. The ranking is based on the stars of each project in GitHub. %Thus, $z$ is the length of the ranking and $+1$ is a correction to avoid the last project having a relevance of $0$. 

The third task is to access to the NPM and GitHub repositories for retrieving features that characterize each technology. We focus on features that can be used as criteria for decision making, such as: number of developers that maintain the technology, \#daily downloads, or \#dependencies, among others. More than 40 features are taken into account\footnote{The full list of features can be found in the Supplementary Material zip file at https://bit.ly/2w9sOzV.}. It should be noticed that all the technologies are represented with the same set of features. It is up to the machine learning algorithm to decide which features are relevant. Our rationale is that, if a technology $T$ was chosen in a project (over other available options), then there should be a criterion over some technology features, upon which $T$ was considered more relevant than the other options. Thus, being able to learn the selection criteria depends on the features collected from the technologies.

\subsubsection{Create dataset}
This step assembles the dataset to be used for building the machine learning (ranking) model. We refer to this dataset as the \emph{training dataset} (Figure  \ref{fig:ApproachOverview}), and it contains a set of \emph{training instances}. A given training instance captures a pair of technologies. %The step performs three tasks.

Initially, a \emph{training ranking} is computed for each technology according to its CDSel value. In our example, \emph{moment} will be ranked first since its CDSel value is higher than the CDSel value of \emph{date-fns} and \emph{momentjs}. Then, each technology is represented as a feature vector $[FT_{i1}, FT_{i2}, ..., FT_{in}]$ where $FT$ is a particular feature and $n$ is the total number of features. Each vector is normalized via feature scaling \cite{jain2011min}, so that its values are between $0$ and $1$. For example, in Figure \ref{fig:dataGeneration} vector (1, 10, 5) for \emph{moment} becomes (1, 0.77, 1) after normalization.%For example, lets assume that the vectors have the characteristics $(has\_keywords, \#releases\_last\_year,\#maintainers)$ where the \emph{moment} and \emph{date-fns} vectors are $(1,10,5)$ and $(0,13,2)$, respectively. Then, after normalization, the vector will be  $(1,0,1)$ and $(0,1,0)$.

At last, a set of \emph{training instances} is created. %A training instance captures a pair of technologies represented as a normalized feature vector. 
Specifically, for each possible pair of technologies in a \emph{training ranking}, a vector is created by concatenation of their normalized feature vectors. A label of $1$ is added to this vector when the first technology is more relevant than the second one, or $0$ otherwise. In our example, the pair \emph{(moment,date-fns)} is labeled to $1$ because \emph{moment} is ranked before \emph{date-fns}. %\adp{la ultima parte/tabla del ejemplo es un poco criptica, yo mejoraria un poco esa parte de la figura}

\subsubsection{Build ML model}
%This step trains the machine learning model using the set of \emph{training instances} created before. 
%Given the nature of the problem of technology selection, 
This step applies a %supervised pairwise 
``learning-to-rank" \cite{li2011learning} technique on the training instances of the dataset. %The output is a model that, given two technologies, predicts an order for them according to their relevance.
A pairwise supervised variant is used, since the order of two given technologies is not affected by  other technologies in the list. For example, \emph{moment} is more relevant than \emph{date-fns} regardless of other alternative technologies. Among the various learning-to-rank algorithms reported in the literature \cite{li2008mcrank}, we chose GBRank \cite{zheng2008general} 
due to its effectiveness in Web search ranking \cite{bai2010cross,kanungo2009web,bian2008finding}. GBRank is based on a gradient boosting method for minimizing wrong preference predictions. %We choose GBRank because its effectiveness has been previously reported in the context of web search ranking \cite{bai2010cross,kanungo2009web,bian2008finding}. 

\subsubsection{Rank technologies}
Given two technologies, the ML model (above) is able to predict an order for them according to their relevance. Finally, this step takes all possible pairs of technologies (from the \emph{Merge technology results step}) and runs them through the ML model, in order to generate a ranking. The pairs are filtered according to the application context defined by the JS developer. Those technologies being most relevant (i.e., popular) should be ranked first by the ML model.%\adp{En el ejemplo de Fig. 3 esta faltando la "salida" del ST-Rank, o sea, el ultimo step}

%Finally, given a list of candidate technologies and using the model trained previously, this step order the technologies by ranking first the most relevant for developers taking into account the intended application context. 

\section{Evaluation}\label{sec:evaluation}

We performed an empirical evaluation of the \emph{ST-Retrieval} and \emph{ST-Rank} components using datasets sampled from JS projects. The following research questions were addressed:
\begin{itemize}
    \item \textbf{RQ\#1}: How does the performance of \emph{ST-Retrieval} compare to that of existing search engines? 
    \item \textbf{RQ\#2}: What is the quality of the rankings proposed by \emph{ST-Rank} with respect to the rankings from NPM and NPMCompare?
\end{itemize}
The experiments to answer RQ\#1 and RQ\#2 are reported in sub-sections \ref{subsec:evaluationSTRetrieval} and \ref{subsec:evaluationSTRank} respectively. %For each experiment we describe the experimental design, hypotheses, operation and results.
%Finally, we discuss the threats to validity (Section \ref{subsec:Threats-to-Validity}) of both experiments.  

\subsection{Evaluation of ST-Retrieval}\label{subsec:evaluationSTRetrieval}
%In this section, we present an experiment to answer RQ\#1. 
The main goal is to determine whether the performance of \emph{ST-Retrieval} for retrieving a set of JS technologies is better than the performance of \emph{DS} and \emph{GP} search engines.

\begin{table}
	\caption{Reference Queries. \label{tab:queries}}
	\begin{centering}
	\begin{tabular}{l}
		\toprule 
		Queries \\
		\midrule
		check valid email address\\
		download web videos\\
		send sms\\
		quick sort algorithm\\
		filter adult content images\\
		user authentication\\
		extract barcode from image\\
		convert data formats\\
		download free music\\
		convert typewritten image to text\\
		sentiment analysis\\
		third party authentication\\
		convert text to speech\\
		calculate word similarity\\
		translate english to spanish\\
		credit card validation\\
		health tracker\\
		captcha authentication\\
		detect text language\\
		rank aggregation algorithms\\
		mobile app framework\\
		DOM manipulation utils\\
		lightweight 3D graphic library\\
		mathematical functions\\
		scraper\\
		\bottomrule
	\end{tabular}
	\par \end{centering}
\end{table}

\subsubsection{Experimental Design and Operation\label{subsec:Experiment-Design}}
%As we stated in Section \ref{sec:stretrieval}, 
We used NPM as the basis for the evaluation, as NPM is the de-facto package repository for JS technologies. %NPM has more than half a million software packages and has had an exponential growth in recent years.  %\footnote{http://www.modulecounts.com/}.
First, we downloaded the technology registry
from the NPM repository. With this information, we built the repository of technologies (Figure \ref{fig:ApproachOverview}) to a given date (08/28/17), as described in sub-section \ref{subsec:Technology-Extraction}. 
%Then, we interview two senior developers in JS technologies. 
Then, we asked two senior JS developers to record any search (i.e., queries) for technologies in NPM that they would make, during 2 weeks, in their normal projects. After filtering some of their search results (in order to remove very similar queries), we obtained a \emph{reference set} of 25
queries (Table \ref{tab:queries}) representing a variety
of technological needs. We ran \emph{ST-Retrieval} 25 times on this set (once for each query) and stored the lists of outputted technologies. The length of the queries was selected as a normal distribution over the most usual length \cite{taghavi2012analysis}. 
%As we stated in Section \ref{subsection:processQuery} 
The search engines were: NPM, NPMSearch, Google, and Bing. 
%At last, different search engines were selected to carry out the queries, namely: npmjs.com (type i-a) \textendash{} referred to as \emph{Npmjs}, npmsearch.com (type i-b) \textendash{} referred to as \emph{Npmsearch}, google.com (type ii) \textendash{} referred to as \emph{Google}, and bing.com (type ii) \textendash{} referred to as \emph{Bing}.
We also ran a retrieval
effectiveness test for search engines \cite{lewandowski2015evaluating}.
%We used the sample of queries described in Table \ref{tab:Queries-used-and}.

During the \emph{Process query} step, we only took into account 
the first 20 documents from the list of results. This was because users searching the Web (e.g., using Google) are very likely to consider only
the first results \cite{hochstotter2009users}. %For instance, in \emph{Google} users generally browse at most two pages of results (i.e. around 20 results). 
These documents were recorded in order each time we ran the step. 
As part of the \emph{Extract technologies} step, we stored all the technologies identified. At the end, we retrieved a total of 2760 JS technologies (in general, multiple technologies were extracted from the documents returned by \emph{Google} and \emph{Bing}). Then, in order to establish a \emph{baseline}, we asked 12 JS senior developers (different from the ones that produced the reference set of queries) to assess the technology results. Each developer was given all the technologies for a query in random order, and was asked to judge the technology relevance using a binary scale (yes/no). The developers did not know which search engine produced a particular result, and duplicate results
(i.e. results returned by more than one engine) were presented to
them only once. In particular, 11 developers analyzed  the results of 2 queries and 1 developer analyzed the results of 3 queries. On average, each JS developer analyzed 230 technologies.

%\begin{table}
%		\caption{Queries used.\label{tab:Queries-used-and}}
%		\begin{tabular}{ll}
%			\toprule  
%			Queries & Words\\
%			\midrule 
%			send sms & 2\\
%			quick sort algorithm & 3\\
%			filter adult content images & 4\\
%			user authentication & 2\\
%			extract barcode from image & 4\\
%			convert data formats & 3\\
%			convert typewritten image to text & 5\\
%			download free music & 3\\
%			credit card validation & 3\\
%			check valid email address & 4\\
%			detect text language & 3\\
%			sentiment analysis & 2\\
%			third party authentication & 3\\
%			convert text to speech & 4\\
%			calculate word similarity & 3\\
%			translate english to spanish & 4\\
%			download web videos & 3\\
%			health tracker & 2\\
%			rank aggregation algorithms & 3\\
%			mobile app framework & 3\\
%			DOM manipulation utils & 3\\
%			lightweight 3D graphic library & 4\\
%			mathematical functions & 2\\
%			scraper & 1\\
%			captcha authentication & 2\\
%			\bottomrule
%		\end{tabular}
%	%\includegraphics[scale=0.5]{retrieval_files/querydistribution}
%\end{table}

\subsubsection{Metrics\label{subsec:IR-Metrics}}
Information
Retrieval metrics were used for gauging performance\cite{buttcher2016information}, such as: precision, recall, MAP (Mean Average Precision) and DCG (Discounted Cumulative Gain). We leveraged on the reference set of queries and the technology baseline described previously.
%IR can be defined as the activity of obtaining information resources relevant to an information need, from a larger collection of information resources.
%In this work, the information resources are the JS technologies, the information needs are the (technology) queries, and the collection of resources are the available JS technologies.

 In our domain, \emph{precision} is the ratio between the number of relevant technologies recovered and the total number of technologies recovered. The closer the precision value is to $1$, the greater the number of relevant technologies recovered with \emph{ST-Retrieval}.\emph{Recall} is the ratio between the number of relevant technologies recovered and the total relevant technologies known. A recall of $1$ indicates that all relevant technologies have been recovered. In addition, \emph{MAP} \cite{caruana2006empirical} measures, for a set of queries, the average %of the average 
 precision for each query. MAP values close to $1$ mean that the relevant technologies are within the top-ranked positions. At last, \emph{DCG} \cite{jarvelin2000ir} measures the quality of rankings, in terms of the utility (gain) of a result based on its relevance and position in the ranking. DCG can be divided by the maximum value taken from among all the queries, and then, the values for each query can be averaged to measure the average quality of the rankings. This variant of DCG is called \emph{Normalized DCG }(\emph{nDCG}).

\subsubsection{Analysis of Results\label{subsec:Results-and-Analysis}}
\emph{Recall} and \emph{Precision} were
calculated in different positions of the ranking in order to establish the
value of the metric in that position. In addition, the reported results
are the arithmetic mean of the values for each query.

\begin{figure*}[h]
	\begin{center}
		\includegraphics[scale=0.32]{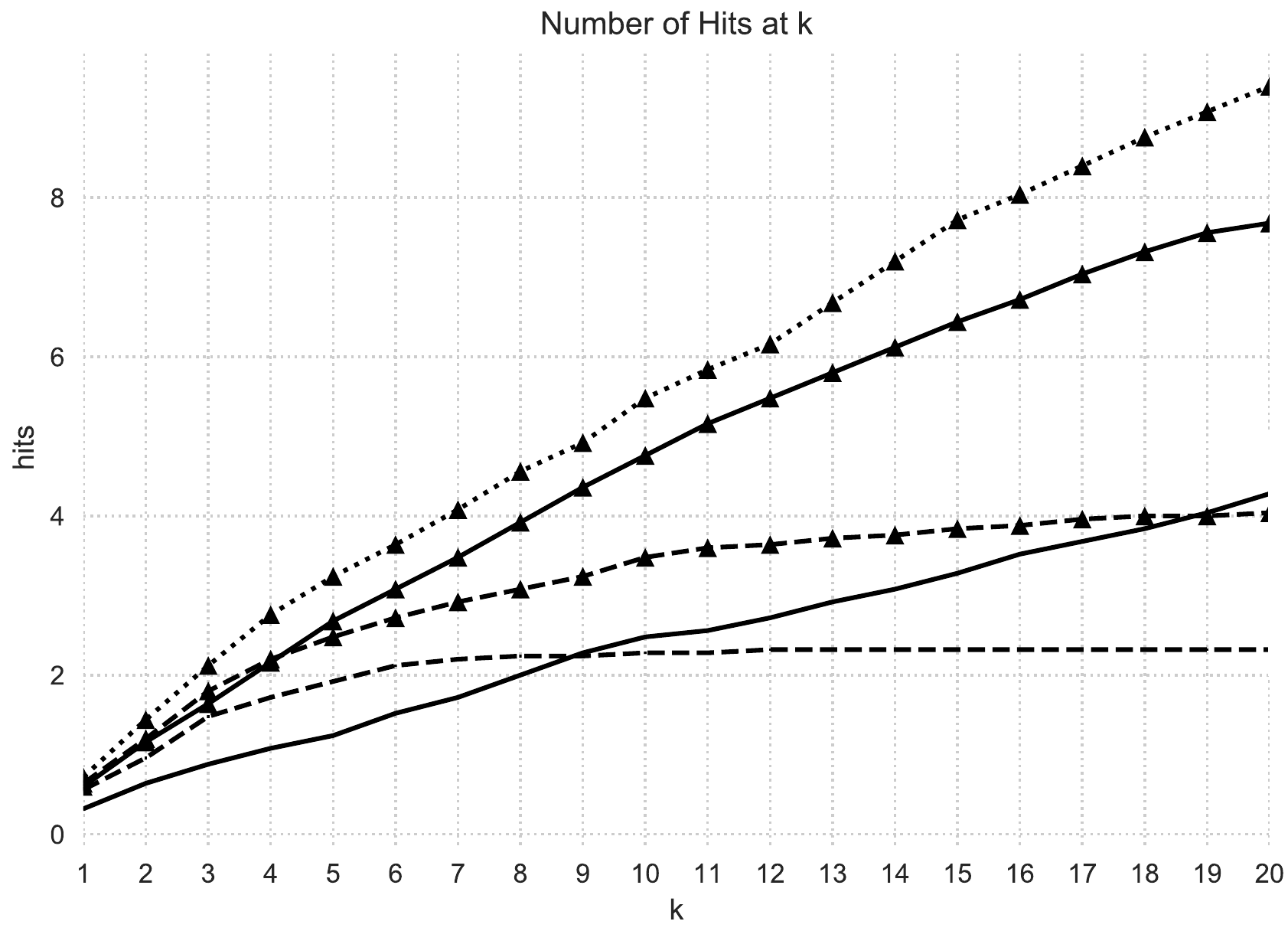}
		\includegraphics[scale=0.32]{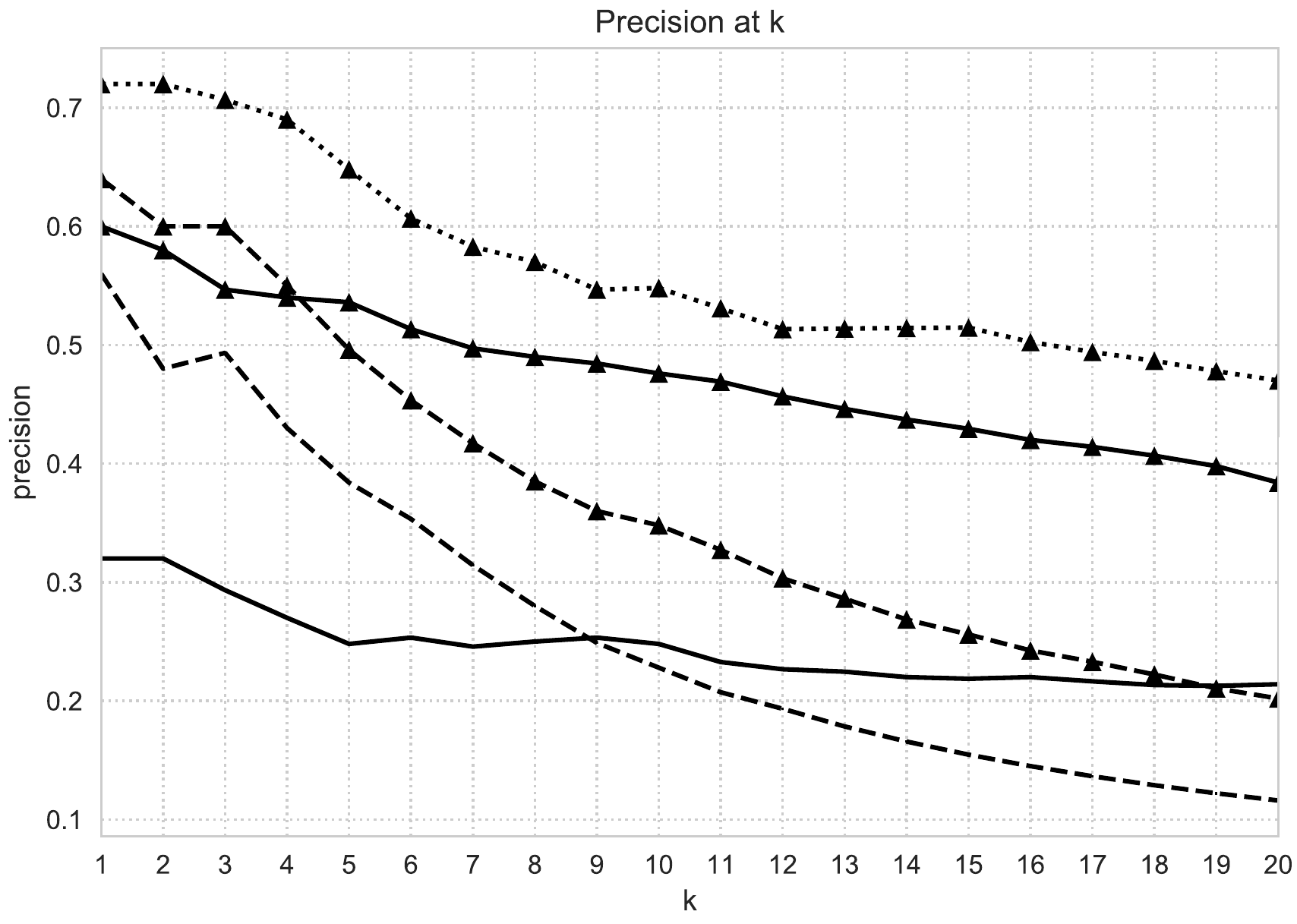}
		\includegraphics[scale=0.32]{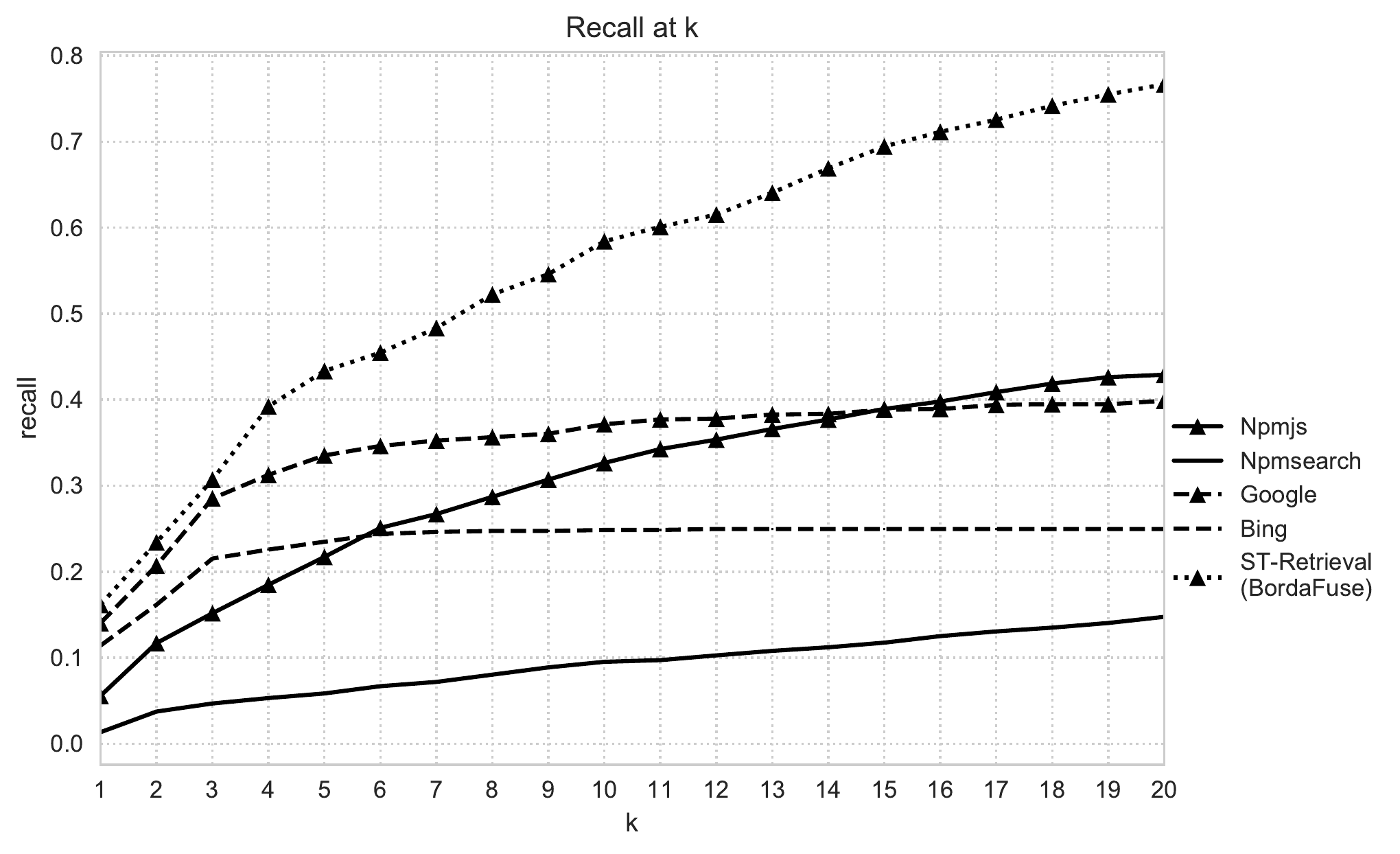}
		%
		%\includegraphics[scale=0.35]{retrieval_files/fmeasurecurveBW}
		%\sv{Podemos quitar F-measure? De esa forma puedo poner los 3 graficos en linea}
	\end{center}
	\caption{Hits, Precision, and Recall at K position (\emph{ST-Retrieval}).\label{fig:Hits,-Precision,-Recall}}
\end{figure*}

Figure \ref{fig:Hits,-Precision,-Recall} shows the results for the four search engines and the Borda Fuse function considering each position ($k$) of the ranking (k
= 1, first position). The first chart shows the average number of
hits (\emph{Hits}). A result is said to be a hit if it is relevant to
the query. The other charts show the results for \emph{Precision},
and \emph{Recall}. As it can be observed, \emph{ST-Retrieval} presents high values in all positions for all the metrics. \emph{Google} and \emph{NPM} behave similarly
to \emph{ST-Retrieval} until around the third position. From that position on, \emph{ST-Retrieval} exhibits a considerable improvement. We argue that this improvement is due to the aggregation of results, and also to the ability of \emph{Borda Fuse} to find the global relevance of a result based on its position in the results relative to the different search engines.  
%If we take Table 2 as an example, it can be clearly seen how the resulting ranking has more relevant results than any individual technique (higher recall) and better results in the first positions (higher precision up to the fourth position). Coming back to the analysis of results, 
The maximum precision obtained by \emph{ST-Retrieval} was 0.72 (8\% increase when compared to the search engines) and the maximum recall was of 0.77 (34\% more than the search engines). This difference in performance is also observed in the nDCG and MAP metrics for ranking quality metrics.

\begin{table*}[h]
	\caption{Metrics for search engines (\emph{ST-Retrieval}).\label{tab:Metrics-for-Search}}
		\begin{tabular}{llllllllllllll}
			\toprule 
			{Method/Metric} & {H@1} & {H@5} & {H@10} & {H@20} & {P@5} & {P@10} & {P@20} & {nD(r=5)} & {nD(r=10)} & {nD(r=20)} & {M(r=5)} & {M(r=10)} & {M(r=20)}\\
			\midrule
			%Search Engine
			%&&&&&&&&&&&&&\\
			%\midrule

			{NPM} & {0.600} & {2.680} & {4.760} & {7.680} & {0.536} & {0.476} & {0.384} & {0.635} & {0.639} & {0.660} & {0.632} & {0.620} & {0.615}\\

			{NPMSearch} & {0.320} & {1.240} & {2.480} & {4.280} & {0.248} & {0.248} & {0.214} & {0.322} & {0.333} & {0.361} & {0.348} & {0.336} & {0.325}\\

			{Google} & {0.640} & {2.480} & {3.480} & {4.040} & {0.496} & {0.348} & {0.202} & {0.720} & {0.736} & {0.750} & {0.712} & {0.704} & {0.695}\\

			{Bing} & {0.560} & {1.920} & {2.280} & {2.320} & {0.384} & {0.228} & {0.116} & {0.662} & {0.675} & {0.677} & {0.638} & {0.639} & {0.639}\\
			
			\midrule
		%	ST-Retrieval
		%	&&&&&&&&&&&&&\\ 
		%	\midrule
			
			{BordaFuse} & \textbf{0.720} & \textbf{3.240} & \textbf{5.480} & \textbf{9.400} & \textbf{0.648} & \textbf{0.548} & \textbf{0.470} & \textbf{0.775} & \textbf{0.770} & \textbf{0.831} & \textbf{0.805} & \textbf{0.776} & \textbf{0.749}\\

		%	{M1} & {0.720} & {3.000} & {5.280} & {9.280} & {0.600} & {0.528} & {0.464} & {0.726} & {0.741} & {0.817} & {0.760} & {0.746} & {0.728}\\
 
		%	{M2} & {0.720} & {3.040} & {5.280} & {9.240} & {0.608} & {0.528} & {0.462} & {0.734} & {0.743} & {0.818} & {0.766} & {0.747} & {0.730}\\

		%	{M3} & {0.680} & {2.920} & {5.240} & {9.280} & {0.584} & {0.524} & {0.464} & {0.701} & {0.724} & {0.803} & {0.737} & {0.725} & {0.710}\\

		%	{M4} & \textbf{0.800} & {2.920} & {5.160} & {9.320} & {0.584} & {0.516} & {0.466} & {0.732} & {0.741} & {0.821} & {0.790} & {0.749} & {0.724}\\
 
		%	{Condorcet} & \textbf{0.800} & {2.880} & {5.20} & {9.360} & {0.576} & {0.520} & {0.468} & {0.722} & {0.741} & {0.820} & {0.779} & {0.740} & {0.723}\tabularnewline
			\bottomrule
		\end{tabular}
\end{table*}
 
Table \ref{tab:Metrics-for-Search}
shows the values for \emph{nDCG} (nD) and \emph{MAP} (M) for ranking sizes of 5, 10, and 20. Table \ref{tab:Metrics-for-Search}
also shows the values of \emph{Hits} (H) and \emph{Precision} (P) in positions 5, 10, and 20, respectively. %In terms of ranking quality, the table shows similar trends in both \emph{nDCG} and \emph{MAP}. 
Interestingly, \emph{Borda Fuse} shows the best performance
along all positions. In the case of the search engines, \emph{NPM} obtained the best results. For this reason, we compared \emph{Borda Fuse} against \emph{NPM}. For instance, \emph{NPM} is outperformed by the \emph{BordaFuse} technique in 19.06\% (in average) for all positions in \emph{Hits}. This means that around position 5 \emph{Borda Fuse} is able to return one more relevant technology than \emph{NPM}. Thus, the precision of \emph{NPM} is
outperformed by \emph{Borda Fuse} by 19.06\% (in average) over all
positions.

Regarding the ranking quality, \emph{Borda Fuse} obtained a \emph{nDCG}
of 22.04\%, 20.5\% and 25.9\% higher than the same metric for \emph{NPM}, for ranking lengths of 5, 10 and 20 respectively. When it comes to \emph{MAP}, the improvements of \emph{Borda Fuse} with respect to
\emph{NPM} were of 27.37\%, 25.16\%, and 21.78\% for ranking lengths
of 5, 10 and 20 respectively.
 The improvement in ranking quality can be related to the improvements in precision and recall. On one hand, by increasing the precision in the top-ranked positions, the relevant results near the top of the ranking increase and so do the values of nDCG and MAP (since both metrics reward results in the top-ranked positions). On the other hand, by increasing recall, the amount of relevant results that add up along the nDCG and MAP calculations also increases.

After observing that the performance values of \emph{ST-Retrieval} were higher than those of the search engines,% no conclusion can be made so far on whether there is a statistical significant difference between them. For this reason, 
we tested the statistical significance of the results using the non-parametric Wilcoxon Signed
Rank Test \cite{lavrenko2017relevance} with a significance level $\alpha=0.05$. 
We stated the null hypothesis ($H1_0$) as: ``The metric values
from \emph{ST-Retrieval} are equal to those
from other search engines". The alternative hypothesis ($H1_1$) states that there is a difference between these metrics. Table \ref{tab:Comparison-on-the} shows the p-values obtained for each metric (treatment) reported in Table
\ref{tab:Metrics-for-Search} after running the tests. With the exception of the number of hits at first position (H@1), all p-value values are less than 0.05. This means that we can reject $H1_0$ for all the metrics (except H@1), and conclude that the differences between \emph{ST-Retrieval} and the existing search engines are statistically significant. Finally, we successfully answer RQ\#1 by saying that \emph{ST-Retrieval} does improve the search results.

\begin{table*}[h]
		\caption{Comparison on the retrieval metrics using Wilcoxon signed rank test (\emph{ST-Retrieval}).\label{tab:Comparison-on-the}}
		\begin{tabular}{llllllllllllll}
			\toprule 
			{ST-Retrieval vs} & {H@1} & {H@5} & {H@10} & {H@20} & {P@5} & {P@10} & {P@20} & {nD(r=5)} & {nD(r=10)} & {nD(r=20)} & {M(r=5)} & {M(r=10)} & {M(r=20)}\tabularnewline
			\midrule
			{NPM} & {.317} & {\textless{}.001} & {\textless{}.001} & {\textless{}.001} & {\textless{}.001} & {\textless{}.001} & {\textless{}.001} & {\textless{}.05} & {\textless{}.05} & {\textless{}.05} & {\textless{}.05} & {\textless{}.001} & {\textless{}.001}\tabularnewline

			{NPMSearch} & {\textless{}.05} & {\textless{}.001} & {\textless{}.001} & {\textless{}.001} & {\textless{}.001} & {\textless{}.001} & {\textless{}.001} & {\textless{}.001} & {\textless{}.001} & {\textless{}.001} & {\textless{}.001} & {\textless{}.001} & {\textless{}.001}\tabularnewline

			{Google} & {.317} & {\textless{}.001} & {\textless{}.001} & {\textless{}.001} & {\textless{}.001} & {\textless{}.001} & {\textless{}.001} & {\textless{}.05} & {\textless{}.05} & {\textless{}.05} & {\textless{}.001} & {\textless{}.001} & {\textless{}.001}\tabularnewline
 
			{Bing} & {.248} & {\textless{}.001} & {\textless{}.001} & {\textless{}.001} & {\textless{}.001} & {\textless{}.001} & {\textless{}.001} & {\textless{}.05} & {\textless{}.001} & {\textless{}.001} & {\textless{}.001} & {\textless{}.001} & {\textless{}.001}\tabularnewline
			\bottomrule
		\end{tabular}
\end{table*}

\subsection{Evaluation of ST-Rank}\label{subsec:evaluationSTRank}
%In this section, we present an experiment to answer RQ\#2. 
The main goal is to determine whether \emph{ST-Rank} improves the rankings generated by NPM. 

\subsubsection{Experimental Design and Operation\label{subsec:Experiment-Design-STRank}}
In order to obtain a ``group of popular projects" for the \emph{Collect data} step, we selected the top-1000 most popular projects according to GitHub, from the repository of technologies (Figure \ref{fig:ApproachOverview}) used in the \emph{ST-Retrieval} experiment. %As it was described in Section \ref{subsec:collectData}, the \emph{Collect data} step uses NPM and NPMCompare to obtain the characteristics and alternatives to each technology. 
NPM and NPMCompare were employed for obtaining features and alternatives for each technology. %We only considered technologies for which at least one alternative was recovered. 
In the \emph{Create dataset} step,
we then created around 250 \emph{training rankings} of between 2 and 6 technologies each. In total, more than 1000 \emph{training instances} were created\footnote{The training dataset can be found in the Supplementary Material zip file at https://bit.ly/2w9sOzV.}. 

To assess the rankings produced by \emph{ST-Rank}, we defined a \emph{test set} by randomly removing 20\% of the \emph{training rankings} (along with their \emph{training instances}). Specifically, we shuffled the order of each \emph{training ranking}. Then, we presented the technologies of each ranking to two senior JS developers and asked them to produce \emph{reference rankings} by sorting the technologies in descending order of relevance\footnote{The reference rankings can be found in the Supplementary Material zip file at https://bit.ly/2w9sOzV.}. The remaining 80\% of the \emph{training instances} were divided to perform a k-fold cross-validation \cite{refaeilzadeh2009cross} with $k=5$, in order to train the ML model and find the best configuration of hyper-parameters for GBRank. Specifically, we used random search \cite{bergstra2012random} to explore the possible values of the hyper-parameters, and the values close to the best random configuration achieved by a grid search \cite{bergstra2011algorithms} were applied. %Figure \ref{fig:roc-auc} shows the results of ROC-AUC for GBRank in the scenarios. 
In the end, GBRank was run with parameters $learning\_rate=0.004$, $max\_depth=50$, $min\_samples\_split=50$, and $min\_samples\_leaf=10$.

For the application contexts, we manually classified the JS technologies based on their execution environment. We considered three possible environments, namely: (i) technologies running in a Web browser (\emph{Web}), (ii) technologies to be used in the Node.js execution environment (\emph{Node}), and (iii) technologies that do not require a specific environment (\emph{No context}). From these application contexts, we configured 5 possible scenarios, as follows: (i) \emph{All} -  the developer makes no distinction among application contexts (i.e. the three execution environments are considered), (ii) \emph{Web} - the developer needs a technology for a Web browser (i.e. \emph{Web} and \emph{No context} are considered), (iii) \emph{Node} - the developer needs a technology for Node.js (\emph{Node} and \emph{No context} are considered), (iv) \emph{OnlyWeb} - the developer needs a technology specifically developed for the Web browser (only \emph{Web} is considered), and (v) \emph{OnlyNode} - the developer needs a technology specifically developed for Node.js (only \emph{Node} is considered). Based on these scenarios, we need \emph{ST-Rank} to classify the inputted technologies and the \emph{training rankings}. Thus, we ran \emph{ST-Rank} 5 times, once for each scenario. 
At last, we compared our results with the default ranking techniques supported by NPM and NPMCompare, which are the Analytic Hierarchy Process (AHP) \cite{saaty2008decision} and the Weighted Average Method (WAM) \cite{perez2000evaluation}, respectively. 

\subsubsection{Metrics}
We use three metrics to evaluate the performance of the rankings, namely: MAP (see Section \ref{subsec:IR-Metrics}), SRCC (Spearman Rank Correlation Coefficient) \cite{zwillinger1999crc}, and MRR (Mean Reciprocal Rank) \cite{chapelle2009expected}. SRCC measures the correlation between the rankings created by \emph{ST-Rank} (and also by AHP and WAM) against the reference rankings created by the senior JS developers. In particular, we calculated SRCC for each ranking and then averaged them to obtain a value that summarizes the ``stability" of each technique. MRR, in turn, evaluates if the  highest-ranked items are relevant. The closer the MRR value is to $1$, the greater the number of relevant technologies in the highest positions.

%\begin{figure}[h]
%	\begin{center}
%		\includegraphics[scale=0.46]{retrieval_files/ROCGBRank}
%	\end{center}
%	\caption{ROC-AUC for GBRank.\label{fig:roc-auc}}
%\end{figure}

%Also, we use ROC-AUC (Area Under the ROC Curve) \cite{caruana2006empirical} to evaluate the GBRank pairwise ranking method and, in this way, determine the setting of the machine learning parameters. As explained previously, the paired learning to rank algorithms reduce the learning problem to a binary classification problem between two elements, where the label indicate the relative order between the elements (that is, 1 if the first element must be ordered first than the second, 0 otherwise). In this sense, adjusting the hyperparameters allows to improve the performance of the classification.  For this goal, a random search \cite{bergstra2012random} was used to explore the possible values of the hyperparameters, and then, the values close to the best random configuration achieved by a grid search \cite{bergstra2011algorithms} were explored. Figure \ref{fig:roc-auc} shows the results of ROC-AUC for GBRank in the scenarios. Specifically, we used $learning\_rate=0.004$, $max\_depth=50$, $min\_samples\_split=50$, and $min\_samples\_leaf=10$.

\subsubsection{Analysis of Results\label{subsec:Results-and-Analysis-STRank}}

\begin{table*}[h]
\caption{Ranking results for different scenarios (\emph{ST-Rank}).}\label{table:rankingResults}
\begin{centering}
\setlength{\tabcolsep}{2pt}
\begin{tabular}{lllllllllllllllllllll}
\toprule
 &\multicolumn{4}{c}{All} & \multicolumn{4}{c}{Web} & \multicolumn{4}{c}{Node} & \multicolumn{4}{c}{OnlyWeb} & \multicolumn{4}{c}{OnlyNode}\tabularnewline
\midrule
Technique & M@3 & M@5 & SRCC & MRR & M@3 & M@5 & SRCC & MRR & M@3 & M@5 & SRCC & MRR & M@3 & M@5 & SRCC & MRR & M@3 & M@5 & SRCC & MRR\tabularnewline
\midrule 
AHP& 0.517 & 0.541 & 0.454 & 0.654 & 0.756 & 0.762 & 0.731 & 0.865 & 0.541 & 0.561 & 0.365 & 0.667 & 0.785 & 0.725 & 0.621 & 0.880 & 0.461 & 0.478 & 0.310 & 0.672\tabularnewline
WAM& 0.813 & 0.798 & 0.656 & 0.863 & 0.744 & 0.729 & 0.669 & 0.836 & 0.793 & 0.806 & 0.668 & 0.869 & 0.928 & 0.928 & 0.878 & 0.952 & 0.822 & 0.796 & 0.750 & 0.883\tabularnewline
GBRank& \textbf{0.925} & \textbf{0.914} & \textbf{0.788} & \textbf{0.915} & \textbf{0.910} & \textbf{0.864} & \textbf{0.835} & \textbf{0.942} & \textbf{0.874} & \textbf{0.889} & \textbf{0.886} & \textbf{0.932} & \textbf{0.952} & \textbf{0.962} & \textbf{0.950} & \textbf{0.964} & \textbf{0.933} & \textbf{0.909} & \textbf{0.907} & \textbf{0.966}\tabularnewline
\bottomrule
\end{tabular}
\par\end{centering}
\end{table*}

%In this section we present the results of the experiment in which we compare the rankings generated by our approach with the ones of AHP and WAM. 
Table \ref{table:rankingResults} shows the metric values for the scenarios. In the case of MAP, we computed it for different ranking lengths (M@3 and M@5 means MAP considers a ranking of 3 and 5 technologies respectively). The values in bold in the columns correspond to the best value for the metric along that column. As it can be observed, GBRank shows the best results for all the metrics. For example, for the \emph{All} scenario, when considering MAP, GBRank outperforms AHP and WAM by around 44\% and 12\% respectively. Similarly, in the case SRCC, GBRank improves the values of AHP and WAM by around 42\% and 17\%. MRR shows smaller improvements of GBRank, 28\% and 6\% for AHP and WAM respectively. The differences between the techniques are similar for the other scenarios. On average, the improvements of \emph{ST-Rank} over AHP and WAM are of 10\%, 20\% at least, and 5\% for MAP, SRCC, and MRR, respectively. The smallest improvement was that of MRR, although this result was expected. MRR only takes into account the first element of the ranking. That is, if the first element of the reference ranking is in the first position of the ranking being evaluated, this ranking has a maximum value of MRR (although the other elements  are disordered). However, in the problem of ranking of technologies, it is important that the other elements are also ordered. If for some reason (e.g., due to a technical limitation) the first technology cannot be used, the rest of the technologies should be as orderly as possible. This might happen if the JS developer needs to develop a proof-of-concept, and she should be able to find the right technology as quickly as possible.%, since proof of concept can take considerable effort.

\begin{table}[h]

\caption{Results of Wilcoxon signed rank test (\emph{ST-Rank}).\label{tab:Significancia-estad=0000EDstica-de-st-rank}}
\begin{centering}
{\scriptsize{}}%
\begin{tabular}{ccccc}
\hline 
{ST-Rank vs} & {M@3} & {M@5} & {SRCC} & {MRR}\tabularnewline
\hline 
{AHP} & {4.41$e^{-12}$} & {1.31$e^{-13}$} & {2.35$e^{-16}$} & {4.27$e^{-12}$}\tabularnewline
{WAM} & {5.52$e^{-05}$} & {1.01$e^{-06}$} & {1.67$e^{-10}$} & {5.79$e^{-04}$}\tabularnewline
\hline 
\end{tabular}%\sv{Como se haya calculo esta tabla es un misterio. El facha tampoco se acuerda. Pero IMHO, todas estas tablas hay que dejarlas. Es preferible que se vea que se hizo el trabajo de tests estadisticos a que falte.}
\par\end{centering}
\end{table}

To analyze if the results are statistically significant, we tested the results using the non-parametric Wilcoxon Signed
Rank Test with a significance level $\alpha=0.05$. 
We stated the null hypothesis ($H2_0$) as: \textquotedblleft The metric values from \emph{ST-Rank} are equal to those
provided by the other techniques (i.e. AHP and WAM)\textquotedblright. The alternative hypothesis ($H2_1$) states that there is a difference between these metrics. Table \ref{tab:Significancia-estad=0000EDstica-de-st-rank} shows the p-values obtained for each metric reported in Table
\ref{table:rankingResults} after running the tests. As it can be seen,
 all p-value values are less than 0.05. This means that we can reject $H2_0$ for all the metrics. Thus, we can conclude that the differences between \emph{ST-Rank} and AHP and WAM are statistically significant. In summary, we successfully answer RQ\#2 by saying that the ranking proposed by \emph{ST-Rank} are better than the ranking generated by the existing engines.

\subsection{Threats to Validity\label{subsec:Threats-to-Validity}}

A threat to construct validity has to do with the queries (reference set) and technology searches (baseline) used in the experiments. We tried to rely on queries and searches being representative of real-world JS development. To this end, we extracted a dataset from the NPM repository using the public JS package registry. 
%In addition, we carefully followed best practices for evaluating recovery effectiveness of search engines to decrease conclusion instability \cite{lewandowski2015evaluating}. 
While we used only 25 queries, they returned 2760 JS technologies that were manually analyzed. Since the analysis of query results from search
engines takes a substantial amount of time from experts, we
preferred not to do a detailed query analysis and leave this for future work. 
%One potential research direction is, therefore, building up an online search engine for sharing results with the community and obtain most accurate feedback.

A threat to internal validity is the usage of Borda Fuse to order the list of technologies in \emph{ST-Retrieval}, which might have biased the results, and also affected the outputs of \emph{ST-Rank}. Alternative aggregation methods could return different orderings and should be explored in future works. Similarly, for \emph{ST-Rank} we used a particular learning-to-rank technique,  but other techniques could have led to different results. 

Another threat is that the training rankings \emph{ST-Rank} were based on the CDSel metric. This metric was chosen as a proxy for the popularity of JS packages, but it might not correctly represent the relevance of the packages for the community. Thus, a better validation of this metrics should be pursued. 

To mitigate threats to external validity, we considered queries
with different sizes, purposes and domains. However, our dataset might not be representative of all kinds of JS projects, and further experimentation and surveys of JS projects are necessary. Furthermore, the off-line evaluation showed that the computations of \emph{ST-Retrieval} sometimes take a considerable time, which
might be a problem in an online search engine. We have not consider yet the computation time of the search engines as a comparison factor.

\section{Related Work}\label{sec:relatedWork}

Several approaches have been developed to support selection of software technologies \cite{grande2014framework}. In general, these approaches are based on creating a list of technologies that are compared and presented to developers, so they can decide which ones to apply to their projects. Some works have focused on the evaluation of a set of pre-established candidate technologies and do not deal with the problem of searching/retrieving the technologies from (Web) repositories. For example, given a set of predefined candidates, Ernst et al. \cite{ErnstKazmanBianco2019} proposes a scorecard to help developers to select a given technology. The scorecard is based on performance, maintenance, and community criteria.  

Software repositories \cite{Clayton} are one of the main sources for search/retrieval of candidate technologies. However, existing repositories have not been very successful for this task, despite improvements in their underlying technology, such as the Web \cite{Clayton}. One of the reasons is the performance of the search engines, which sometimes fail to produce the desired results. Based on the above, several works have tried to improve the search offered by the repositories. As far as we know, no works about meta-search targeted to software technologies have been proposed. However, there are a few works that bear similarities with our approach. Agora \cite{seacord1998agora} is a research prototype that intended to replace the idea of software repositories by creating a global database of JavaBeans and CORBA components, which was automatically generated and indexed. However, a major drawback was its reliance on the syntactic interface of JavaBeans and CORBA to carry out its search function. In addition, Agora did not use search engines for software repositories. Another recent work is Dolphin \cite{zhan2016dolphin}, which indexes open-source projects from version control repositories (e.g., OpenHub, SourceForge) and ranks them according to their influence in discussions of forum communities (e.g., StackOverflow, OSChina). Our main difference with this work is that Dolphin searches open-source code from version control repositories, while \emph{ST-Retrieval} focuses on software released and stored in repositories (e.g., NPM). Another difference is that Dolphin does not use general-purpose search engines. 

LibFinder \cite{ouni2017search} is a search-based recommendation system for Java that uses multi-objective optimization to recommend software libraries, that combine GitHub and Maven repositories. However, it does not allow user queries to find technologies. Instead, it is based on the source code to recommend libraries that can replace pieces of source code made by hand. LibFinder does not use general-purpose search engines. Soliman et. al. \cite{soliman2017developing} developed an approach to retrieve architectural design decisions and solution alternatives. The approach uses StackOverflow as an example of an online repository of architecture knowledge. However, this approach is based on a mapping between text and a ``de-facto" ontology, and its applicability to other contexts is yet to be determined.

Regarding the order in which the technologies are created, there are works that address the problem from different perspectives \cite{basili1991support,birk1997modelling,klein2015design}. However, most of these works create ranking strategies manually, based on specific candidate characteristics. For example, Franch and Carvallo \cite{franch2002quality} propose the adoption of a structured quality model to evaluate software packages. This model provides a taxonomy of software quality characteristics and metrics to calculate its value for a given domain. Jadhav et. al \cite{jadhav2011framework} classifies package ranking strategies into two groups, those using AHP and those using WAM. Then, an expert system is used. This approach is difficult to implement, since it depends on experts to construct rules in a manual fashion. Grande et al. \cite{grande2014framework} define the selection problem as a multi-objective optimization problem and develop a framework for its treatment. They apply genetic algorithms to solve the multi-objective problem. One limitation of this approach is the creation and maintenance of the repository of technologies and their characteristics.

\section{Conclusions}\label{sec:Conclusiones}

%The selection of software technologies is an important aspect for the success of software development. %The problems that arise during the selection of technologies can have negative repercussions on the project. In this paper, we address two problems i) the lack of precision in the search engines available in Web software repositories and ii) the difficulty in evaluating candidate technologies. 
In this article %To face these challenges, 
we proposed an approach targeted to JS technologies based on two complementary phases, called \emph{ST-Retrieval} and \emph{ST-Rank}. \emph{ST-Retrieval} helps developers to search and retrieve JS technologies using a meta-search strategy. \emph{ST-Rank} assists developers in ranking the candidate JS technologies identified by the previous phase. To achieve this goal, \emph{ST-Rank} uses a machine learning strategy that learns how to create rankings of technologies from previous choices made by the development community in popular open-source projects.

An initial evaluation of the approach using the NPM repository showed satisfactory results. In the case of \emph{ST-Retrieval}, precision and recall values of around 80\% were obtained, improving other search engines by approximately 20\%. Regarding \emph{ST-Rank}, improvements of around 5\% for MRR, 10\% for MAP, and 20\% for SRCC, were obtained in the ranking metrics. Despite these results, the approach still presents some drawbacks. First, the technology extraction function using string-matching in \emph{ST-Retrieval} can be expensive in computational terms. This might limit the application of the approach for online searching. For this reason, we will analyze the possibility of building an extraction function using other NER techniques and exploring their influence on the \emph{ST-Retrieval} results. Second the CDSel metric to measure the degree of selection of a technology in the community  in \emph{ST-Rank} still needs further validation. % to assert that it correctly represents good technological decisions. 
Although reasonable results were obtained during the evaluations with CDSel using 1000 projects, considering alternative parameters in this metric can have an impact on the quality of the rankings.

As future work, in the case of \emph{ST-Retrieval}, we will apply aggregation functions other than Borda Fuse. In the case of \emph{ST-Rank}, we will explore alternatives to GBRank for the learning-to-rank task. In addition, we plan to conduct a study with subjects (JS developers as users of our approach) in order to corroborate our findings. Further work can extend our approach to  technology repositories for other programming languages (e.g., Ruby, Python or Java, among others). Finally, following ideas of XAI (eXplainable AI) \cite{gunning2017explainable}, it would be interesting to add support for generating explanations of the rankings, so that the comparison strategies are easier to understand by developers.

\balance
\bibliography{main}

\end{document}